\documentclass{aip-cp}

\usepackage[numbers]{natbib}
\usepackage{rotating}
\usepackage{epsfig,graphicx,pst-all}
\usepackage{amssymb}
\usepackage{mathrsfs}
\usepackage{esvect}
\usepackage{url}
\usepackage{caption}

\begin{document}

\title{Role of spd- Continuum Components in the Halo Nucleus $^6$He}

\author[aff1,aff2]{Jagjit Singh\corref{cor1}}

\affil[aff1]{Dipartimento di Fisica e Astronomia ``G.Galilei'', via Marzolo 8,I-35131 Padova, Italy}
\affil[aff2]{INFN-Sezione di Padova, via Marzolo 8, I-35131 Padova, Italy.}
\corresp[cor1]{Corresponding author: jsingh@pd.infn.it}

\maketitle

\begin{abstract}
Role of different continuum components in the weakly bound nucleus $^6$He is studied by coupling 
unbound spd-waves of $^5$He by using simple pairing contact-delta interaction.
The results show that $^6$He ground state $0^+$ displays collective nature by taking contribution from five
different oscillating continuum states that sum up to give an exponentially decaying bound wavefunction
emerging from five different possible configurations
i.e. $(s_{1/2})^2$, $(p_{1/2})^2$, $(p_{3/2})^2$, $(d_{3/2})^2$ and $(d_{5/2})^2$.
The ground state properties of $^6$He has been calculated.
\end{abstract}

\section{INTRODUCTION}
The nucleus $^6$He is a typical \textquotedblleft neutron skin or halo\textquotedblright \cite{TANI} 
with a very low neutron pair separation energy ($0.975$ MeV) \cite{AJZEN}.
Motivated by the recent experimental measurements at GANIL \cite{MOUG} on continuum resonances in $^6$He,
recently we have developed a simple theoretical model \cite{FORT} to study the weakly bound ground state and 
low-lying continuum states of $^6$He by coupling two unbound p-waves of $^5$He. 
In the present study we have extended the model space with 
inclusion of sd- continuum waves of $^5$He. The large basis set of these spd- continuum wavefunctions are used 
to construct the two-particle $^6$He ground state (g.s.) $0^+$ emerging from five different possible configurations
i.e. $(s_{1/2})^2$, $(p_{1/2})^2$, $(p_{3/2})^2$, $(d_{3/2})^2$ and $(d_{5/2})^2$. The simple pairing 
contact-delta interaction is used and pairing strength is adjusted to reproduce the bound ground state of 
$^6$He. The extension of model space is a computationally challenging problem.
The main aim is to show how an extension of theoretical concepts related to residual interactions, 
namely a contact delta pairing interaction, naturally explain the stable character of the bound states of
Borromean nuclei, such as $^6$He and simultaneously account for some of the resonant structures seen in the 
low-lying energy continuum.
\section{SPECTRUM OF $^5$He}
\begin{figure}[!t]
\includegraphics[width=0.78\textwidth]{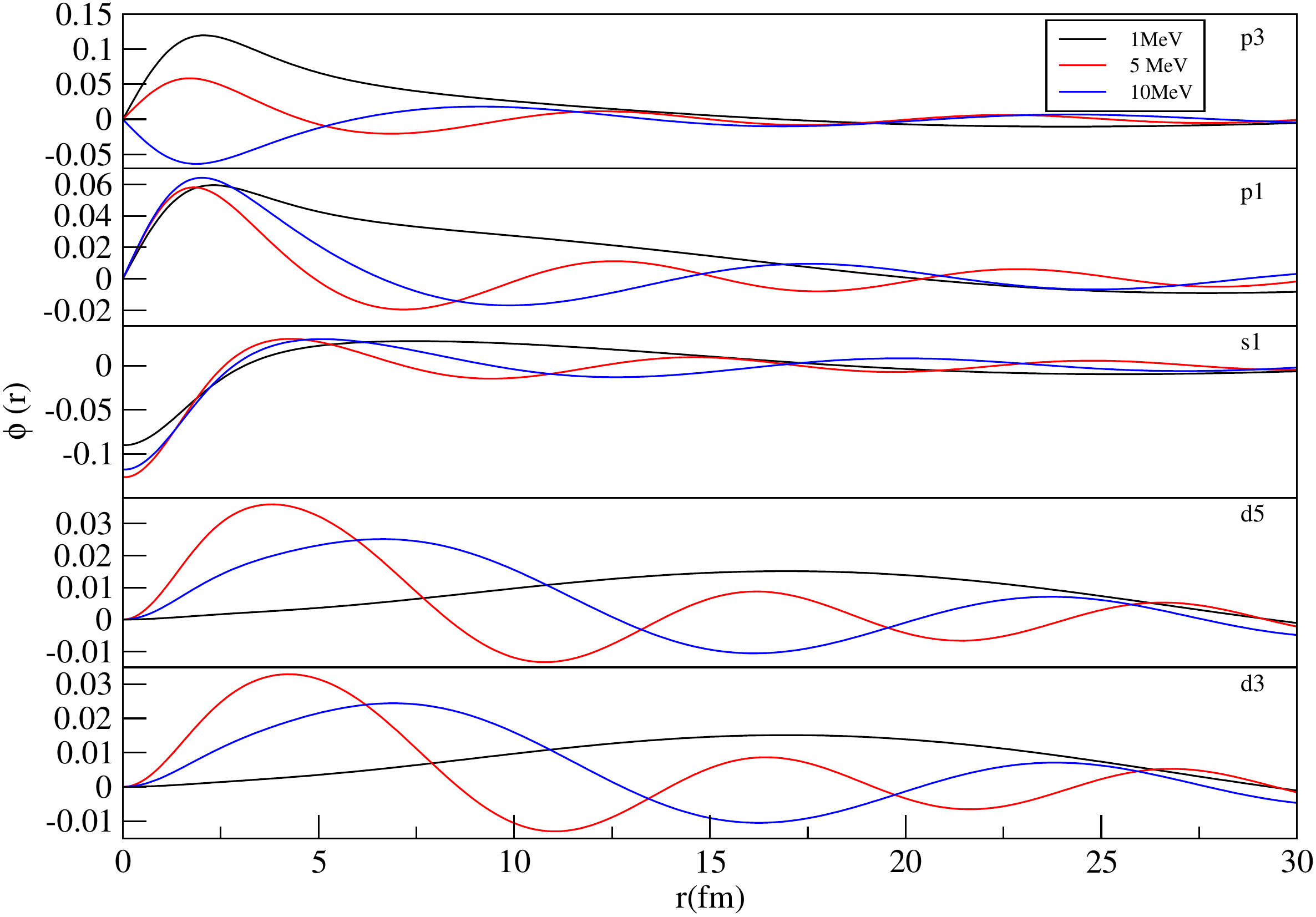}
\caption{(Color online) $^5$He continuum waves as a function of radial variable for continuum energies 1, 5 and 
10 MeV respectively}
\label{5He}
\end{figure}

The unbound nucleus $^5$He can be described as an inert $^4$He core with an unbound neutron moving in $p$ 
doublet or $d$ doublet or $s$ singlet in simple independent-particle shell model picture. 
These doublets are further splitted by spin-orbit interaction. Experimentally only the $p_{3/2}$ and $p_{1/2}$ 
resonances are confirmed at $0.789$ and $1.27$ MeV above the neutron separation threshold and 
their widths are quoted as $0.648$ MeV and $5.57$ MeV respectively \cite{TUNL}. 
Theoretically in order to extend the model space we have also included the $sd-$shell in picture.
The continuum monopole ($\ell=0$), dipole ($\ell=1$) and quadrupole ($\ell=2$) scattering single particle 
states ($E_C>0, k> 0$) of $^5$He (see Fig-(\ref{5He})) are generated with Woods-Saxon (WS) potential given by
\begin{equation}
 V_{WS}=\left[ V_{0} + V_{ls}r_{0}^2(\vv{l}.\vv{s})\frac{1}{r}\frac{d}{dr}\right]\left[1+exp\left(\frac{r_0-r}{a}\right)\right]^{-1} \label{WSP}
 \end{equation}
where $r=r_0A^\frac{1}{3}$. For $^5$He the parameter set used is WS potential depth $V_0=-42.6$ MeV, 
$r_0=1.2$ fm, $a=0.9$ fm and spin-orbit coefficient $V_{ls}=8.5$ MeV.

\subsection{MODEL AND RESULTS}
The simple model with two non-interacting particles in the above single-particle levels of $^5$He produces
different parity states (see TABLE-$1$) when two neutrons are placed in five different $s_{1/2}$, $p_{1/2}$, $p_{3/2}$, 
$d_{3/2}$ and $d_{5/2}$ unbound orbits. Namely five configurations $(s_{1/2})^2$, $(p_{1/2})^2$, $(p_{3/2})^2$, $(d_{3/2})^2$ 
and $(d_{5/2})^2$ couple to $J=0^+$.

\begin{table}[h]
\centering
\caption{Possible configurations of $^{6}$He arising from two neutrons in s-, p- and d-orbitals}
\centering
\vspace*{0.3cm}
\begin{tabular}{llllll}
\hline
\multicolumn{1}{}{}          & $s_{1/2}$ & $p_{1/2}$   & $p_{3/2}$   & $d_{3/2}$               & $d_{5/2}$                \\ \hline
\multicolumn{1}{l}{$s_{1/2}$} & $0^+$     & $0^- , 1^-$ & $1^- , 2^-$ & $1^+ , 2^+$             & $2^+ , 3^+$             \\ 
\multicolumn{1}{l}{$p_{1/2}$} &           & $0^+$       & $1^+ , 2^+$ & $1^- , 2^-$             & $2^- , 3^-$             \\ 
\multicolumn{1}{l}{$p_{3/2}$} &           &             & $0^+ , 2^+$ & $0^- , 1^- , 2^- , 3^-$ & $1^- , 2^- , 3^- , 4^-$ \\ 
\multicolumn{1}{l}{$d_{3/2}$} &           &             &             & $0^+ , 2^+$             & $1^+ , 2^+ , 3^+ , 4^+$ \\ 
\multicolumn{1}{l}{$d_{5/2}$}  &           &             &             &                         & $0^+ , 2^+ , 4^+$       \\ \hline
\end{tabular}
\end{table}
 The two-particle wave functions are constructed by tensor coupling of two continuum states of $^5$He.
The five states discussed above are not discrete, but rather depend on the energies of the continuum orbitals. 
Each single particle continuum wavefunction is given by
\begin{equation}
\phi_{\ell,j,m}(\vec{r},E_C)=\phi_{\ell,j}(r,E_C)[Y_{\ell m_\ell}(\Omega)\times \chi_{1/2,m_s}]^{(j)}_m \label{spwfn} 
\end{equation}
The combined tensor product of these two is given by
\begin{equation}
 \psi_{JM}(\vec{r}_1,\vec{r}_2)=[\phi_{\ell_1,j_1,m_1}(\vec r_1,{E_C}_1) \times \phi_{\ell_2,j_2,m_2}(\vec r_2, {E_C}_2)]^{(J)}_M\label{tpwfn}
\end{equation}
For simplicity an attractive pairing contact delta interaction $(-g\delta(\vec r_1 - \vec r_2))$ is used, because we 
can reach the goal with only one parameter adjustment. 
The continuum single-particle wavefunctions are calculated with energies from 0.0 to 10.0 MeV and 
normalized to a delta for the spd-states of $^5$He on a radial grid which varies from 0.1 fm to 100.0 fm with the 
potential discussed above. The two particle states are formed using mid-point method with an energy spacing of 
2.0, 1.0, 0.5, 0.2 and 0.1 MeV corresponding to block basis dimensions of $N=$5, 10, 20, 50 and 100 respectively and 
the matrix elements of the pairing interaction are calculated.
In Fig-{\ref{eig-vs-basis}, the eigenspectrum for $J=0^+$ case is presented and from figure it is clear 
that with increase in basis dimensions the superflous bound states moves into the continuum.
The coefficient of the $\delta-$contact matrix $(G)$ is adjusted to reproduce the correct ground state energy. 
The actual pairing interaction $g$ is obtained by correcting with a factor which depends on the 
aforementioned spacing between energy states. The correction factor is practically a constant quantity, 
except for the smallest basis. The biggest adopted basis size gives a fairly dense continuum in 
the region of interest.
The radial part of the $S=0$ ground state wavefunction obtained from the diagonalization in the largest basis is 
presented in the left part of Fig-\ref{groundsts}.
It shows a certain degree of collectivity, taking contributions of comparable magnitude from several basis states.
The surface plot shows the exponential behavior typical of a bound state, despite 
being the sum of many products of oscillating wavefunctions.
One can see from the right part of the Fig.-\ref{groundsts} that the square of the amplitudes 
of the $(p_{3/2})^2$ components are dominant summing up to 89.7\%. 
The comparison among present results and earlier studies by T.Myo \cite{MYO} and Hagino \cite{HAG} is presented in 
TABLE-$2$ and from this table it is clear that present results are in good agreement with earlier calculations, which further 
strengthen present results.
In TABLE-$3$, calculated ground state properties are compared with earlier studies\cite{MYO,HAG}, 
where $R_{m}$ is the matter radius,
\begin{equation}
\langle r_{NN}\rangle = \langle \psi_{gs}(\vec{r}_1,\vec{r}_2) |(\vec{r}_1-\vec{r}_2)^2| \psi_{gs}(\vec{r}_1,\vec{r}_2) \rangle
\end{equation}
is the mean square distance between the valence neutrons, and
\begin{equation}
\langle r_{c-NN}\rangle = \langle \psi_{gs}(\vec{r}_1,\vec{r}_2) |(\vec{r}_1+\vec{r}_2)^2/4| \psi_{gs}(\vec{r}_1,\vec{r}_2) \rangle
\end{equation}
is the mean square distance of their centre of mass with respect to the core.
\begin{figure}[!t]
\includegraphics[width=0.68\textwidth]{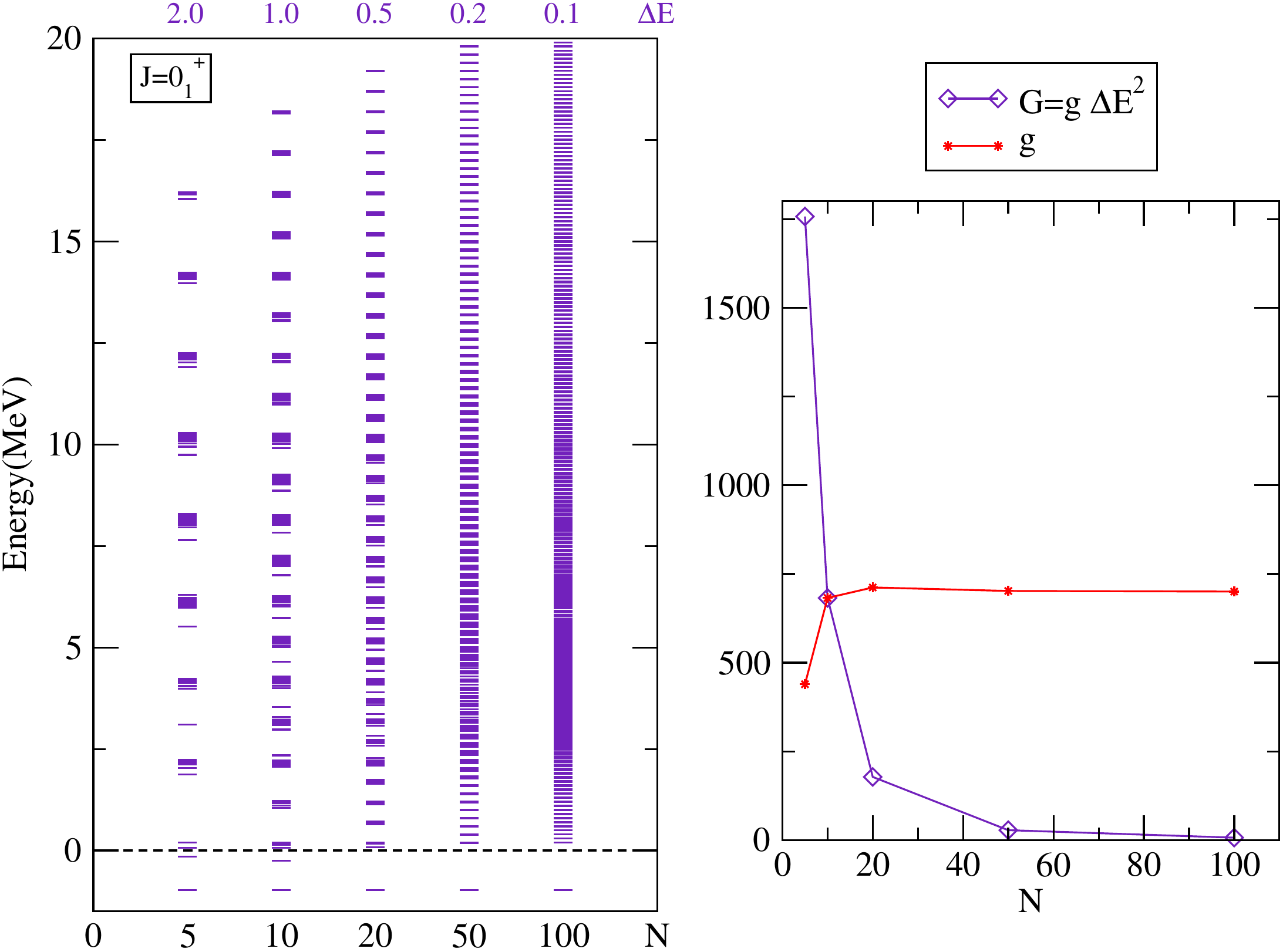}
\caption{(Color online) Eigenspectrum of the interacting two-particle case for $J=0$ for increasing basis dimensions, $N$. 
  The coefficient of the $\delta-$contact matrix, $G$, has been adjusted each time to reproduce the g.s. energy 
  (right). The actual strength of the pairing interaction, $g$, is obtained by correcting with the energy 
  spacing $\Delta E$ and it is practically a constant.}
\label{eig-vs-basis}
\end{figure}

\begin{figure}[!t]
\hspace*{-1.2cm}
\includegraphics[width=0.5\textwidth]{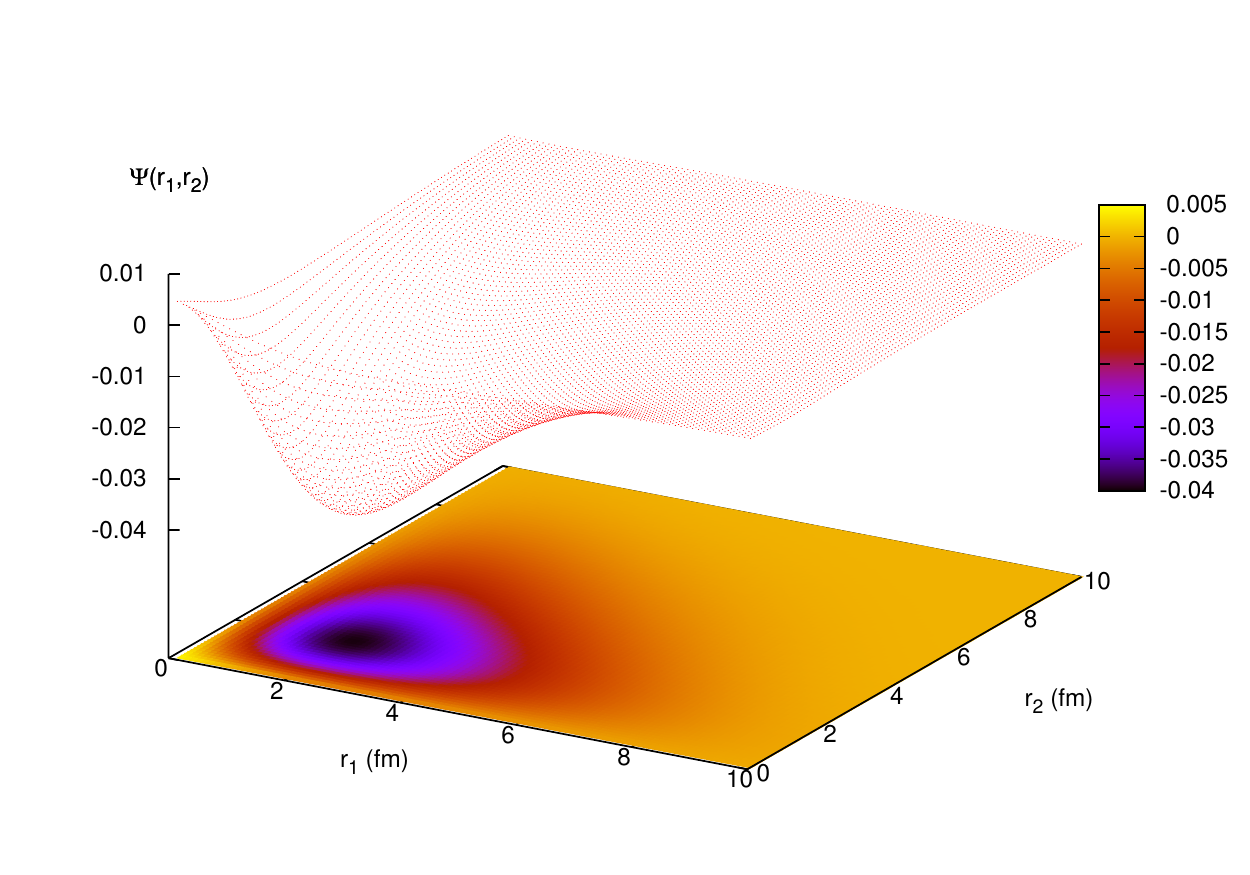}
\hspace*{-0.3cm}
\includegraphics[width=0.5\textwidth]{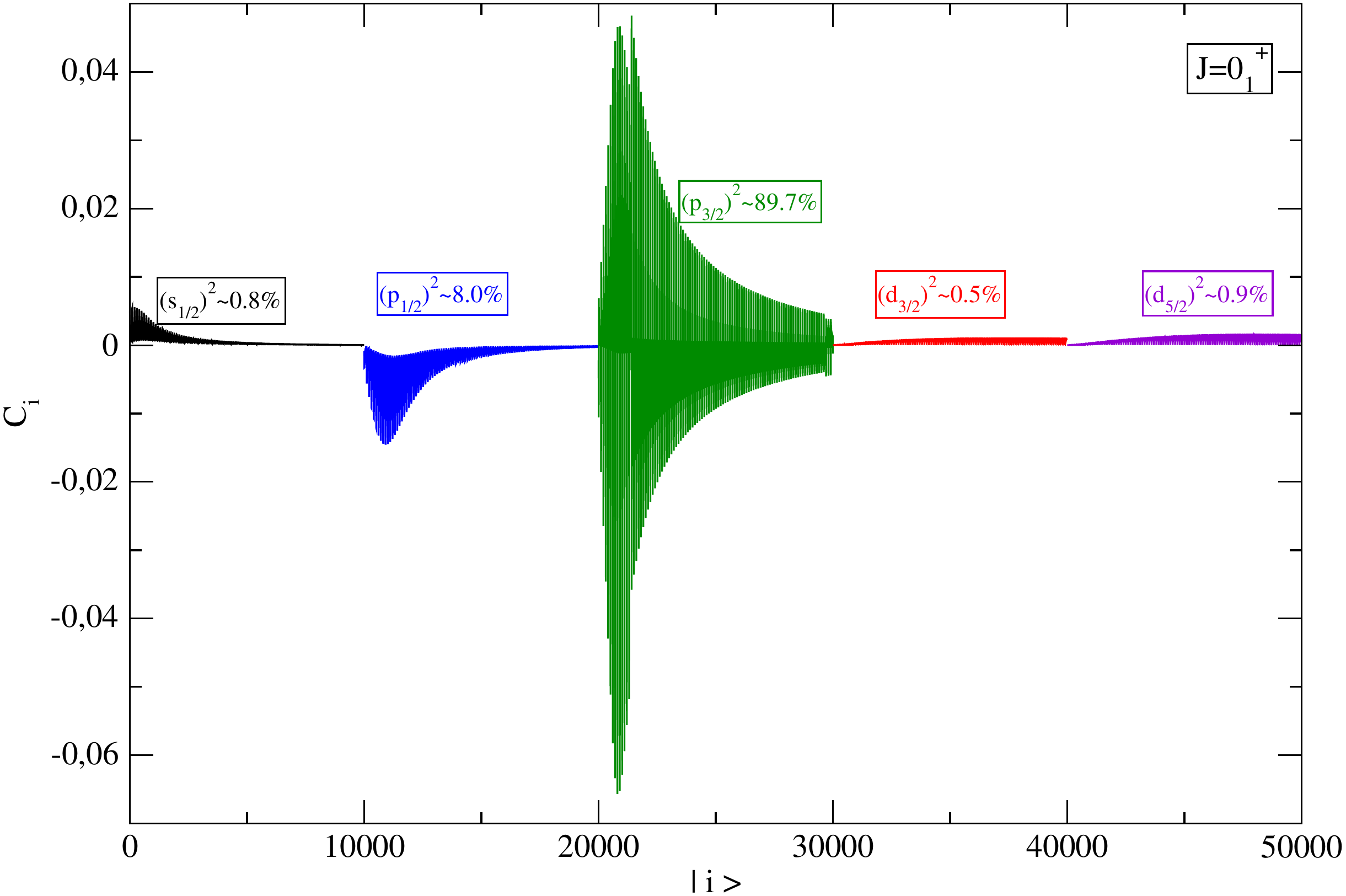}
\caption{(Color online) Ground state wavefunction ($S=0$) for $N=$100 as a function of the coordinates of 
the two neutrons and corresponding contour plot (Left part). Decomposition of the g.s. into the J=0 basis 
(Right part) as a function of an arbitrary basis state label $|i>$: the basis 
is divided into five blocks, $10^4$ $[s_{1/2}\times s_{1/2}]^{(0)}$, $10^4$ $[p_{1/2}\times p_{1/2}]^{(0)}$, $10^4$ $[p_{3/2}\times p_{3/2}]^{(0)}$, $10^4$ $[d_{3/2}\times d_{3/2}]^{(0)}$ and $10^4$ $[d_{5/2}\times d_{5/2}]^{(0)}$ components.. 
The ordering in each block is established by the sequential energies of each pair of continuum s.p. states, 
i.e. $(E_{C_1},E_{C_2}) =$ (0.1, 0.1), (0.1, 0.2), $\dots$ ,(0.1, 10.0), (0.2, 0.1), (0.2, 0.2), $\dots$ 
(10.0, 10.0).} 
\label{groundsts}
\end{figure}
\vspace*{-0.7cm}
\begin{minipage}[b]{.40\textwidth}
  \centering
  \vspace*{0.3cm}
\begin{tabular}{cccc}
\hline
\vspace*{0.3cm}
Config.         &  Present & T.Myo\cite{MYO} & Hagino\cite{HAG} \\ 
\hline
 $(2s_{1/2})^2$  &  0.008           & 0.009 & -- \\
 $(1p_{1/2})^2$  &  0.080          & 0.043  & -- \\
 $(1p_{3/2})^2$  &  0.897           & 0.917 & 0.830\\
 $(1d_{3/2})^2$  &  0.005           & 0.007 &  --\\
 $(1d_{5/2})^2$  &  0.009           & 0.024 &  --\\
\hline
\end{tabular}
  \captionof{table}{Components of the ground state ($0^+_1$) of $^{6}$He}
  \label{components}
\end{minipage}\qquad
\hspace{1.5cm}
\begin{minipage}[b]{.40\textwidth}
  \centering
  \vspace*{0.3cm}
  \begin{tabular}{cccc}
\hline
\vspace*{0.3cm}
         &  Present & T.Myo\cite{MYO} & Hagino\cite{HAG} \\ 
\hline
$R_{m}$   &  $2.37674$           & $2.37$ & ... \\   
 $r_{NN}^2$  &  $28.8404$           & $23.2324$ & $21.3$ \\
 $r_{c-2N}^2$  &  $7.21011$          & $9.9225$  & $13.2$ \\
\hline
\end{tabular}
  \captionof{table}{Radial properties of the ground state of $^{6}$He in units of fm}
  \label{rad}
\end{minipage}
\vspace*{-0.5cm}
\section{CONCLUSIONS}
In the present study I present emergence of bound halo ground state of $^{6}$He from the coupling of five unbound spd- waves in the
continuum of $^{5}$He, due to presence of pairing interaction. Contribution of different configurations has been presented. 
Radial properties of ground state of $^{6}$He are also presented and compared with other calculations.
\vspace*{-0.5cm}
\section{ACKNOWLEDGMENTS}
The useful discussions with L.Fortunato, A.Vitturi, R.Chatterjee and Sukhjeet Singh and financial assistance from 
Fondazione Cassa di Risparmio di Padova e Rovigo (CARIPARO) is gratefully acknowledged.


\nocite{*}
\bibliographystyle{aipnum-cp}%
\bibliography{ref}%

\end{document}